\documentclass[runningheads]{llncs}
\usepackage{cite}
\usepackage{amsmath,amssymb,amsfonts}
\usepackage{algorithmic}
\usepackage{graphicx}
\usepackage{textcomp}
\usepackage[dvipsnames]{xcolor}
\def\BibTeX{{\rm B\kern-.05em{\sc i\kern-.025em b}\kern-.08em
    T\kern-.1667em\lower.7ex\hbox{E}\kern-.125emX}}

\usepackage{hyperref}

\usepackage{newfloat}
\usepackage{syntax}
\usepackage{framed}
\usepackage{etoolbox}
\AtBeginEnvironment{grammar}{\scriptsize}

\usepackage{svg}
\graphicspath{{images/}}
\usepackage[T1]{fontenc}

\usepackage{listings}
\usepackage[linguistics]{forest}
\forestset{
  qtree/.style={
    baseline,
    for tree={
      circle,
      black,
      draw,
      parent anchor=south,
      child anchor=north,
      align=center,
      inner sep=1pt,
    }
  }
}

\usepackage{textcomp}
\usepackage{pgfplots}
\pgfplotsset{width=(\textwidth/2-5pt),compat=1.9}
\usetikzlibrary{patterns}

\usepackage{mathabx}

\usepackage{stfloats}

\usepackage{lipsum}
\usepackage[inline]{enumitem}

\usepackage{subcaption}

\usepackage{multirow}
\usepackage{array}
\newcolumntype{x}[1]{>{\centering\arraybackslash\hspace{0pt}}p{#1}}
\usepackage{dcolumn}

\usepackage{todonotes}

\makeatletter
\newcommand\notsotiny{\@setfontsize\notsotiny\@vipt\@viipt}
\makeatother

\begin{document}

\title{A Query Language for \\
Software Architecture Information \\
(Extended version)}

\author{Joshua Ammermann\inst{1}\orcidID{0000-0001-5533-7274} \and
Sven Jordan\inst{2}\and
Lukas Linsbauer\inst{3}\and
Ina Schaefer\inst{1}\orcidID{0000-0002-7153-761X}}

\authorrunning{J. Ammermann et al.}
%
\institute{
Institute of Information Security and Dependability, \\
Karlsruhe Institute of Technology (KIT), Karlsruhe, Germany \email{\{name\}.\{surname\}@kit.edu} \\
\and
Group IT Solution \& Enterprise Architecture, \\ 
Volkswagen AG, Wolfsburg, Germany \\
\email{sven.jordan@volkswagen.de}
\and
Institute of Software Engineering and Automotive Informatics, \\
Technische Universität Braunschweig, Braunschweig, Germany
}

\maketitle
\begin{abstract}
Software maintenance is an important part of a software system's life cycle.
Maintenance tasks of existing software systems suffer from architecture information that is diverging over time (architectural drift).
The Digital Architecture Twin (DArT) can support software maintenance by providing up-to-date architecture information.
For this, the DArT gathers such information and co-evolves with a software system, enabling continuous reverse engineering. 
But the crucial link for stakeholders to retrieve this information is missing.
To fill this gap, we contribute the Architecture Information Query Language (AIQL), which enables stakeholders to access up-to-date and tailored architecture information.
We derived four application scenarios in the context of continuous reverse engineering.
We showed that the AIQL provides the required functionality to formulate queries for the application scenarios and that the language scales for use with real-world software systems.
In a user study, stakeholders agreed that the language is easy to understand and assessed its value to the specific stakeholder for the application scenarios.

\end{abstract}

\section{Introduction}
As software systems evolve over time, architecture documentation is getting increasingly important. In many cases, the architecture documentation is not kept up-to-date due to cost and time restrictions~\cite{Kazman1999}.
This divergence is called architectural drift and leads to increased maintenance effort~\cite{Taylor2010}.
To mitigate the problem of architectural drift, Software Architecture Recovery (SAR) methods emerged~\cite{Ducasse2009, Garcia2013}.
SAR methods enable the recovery of architecture information but only target isolated views (e.g. structural or behavioral views) of a software system from specific information sources (e.g. source code).
Further, architecture information is only recovered at a single point in time.

To solve all of the aforementioned issues, Jordan et al.~\cite{Jordan2022} recently developed the Digital Architecture Twin (DArT), an architectural data model that co-evolves with a software system. A DArT incorporates several SAR methods to automatically generate up-to-date architecture information from multiple information sources (e.g. source code, build and deployment scripts, and views).
The DArT enables integration into the development process in the form of continuous reverse engineering~\cite{Canfora2011}.
Further, the DArT provides up-to-date, tailored architecture information for stakeholders enabling design decisions on the current software architecture of the software system.
However, the crucial link between stakeholders and extensive architecture data models, i.e., a tailored Architecture Query Language (AQL) for the DArT, is missing.

This ability to query an architecture information base is a challenge in reverse engineering~\cite{Canfora2011}.
The term AQL was proposed by Sartipi~\cite{Sartipi2001} in the context of their reverse engineering approach.
Sartipi's AQL is not suitable for querying the DArT as it is limited in the supported views and constrained to a specific recovery process~\cite{Sartipi2001}, which is too restrictive to use for extensive architecture data models.
The problem that arises with limited views is that accessing the vast amounts of information in architecture data models becomes harder, the fewer views exist in the query language.
In fact, some information may be irretrievable if there is no suitable view to query it.
The AQL by Wang et al.~\cite{Wang2004} is tailored to information exchange between Architecture Description Languages (ADLs), which document software architectures~\cite{Medvidovic1997, Taylor2010}, so the query results are ADL instances themselves~\cite{Wang2004}.
This is inconvenient to stakeholders because for each view a suitable ADL has to be known and used.

To counter those shortcomings of existing AQLs, we propose the Architecture Information Query Language (AIQL)\footnote{This work briefly summarizes the author's master thesis\cite{Ammermann2022} and provides further evaluation regarding the AIQL's scalability and usability.} that is process-agnostic, typed and supports querying architecture information of various software architecture views.
The AIQL is easily extensible to support additional views and has an easy-to-use syntax.
Architecture information can be queried by defining and restricting query templates.
Composition of query templates is a key element of the AIQL, that strongly encourages reuse.
The AIQL provides stakeholders with efficient access to the architecture information in extensive architecture data models.
To demonstrate this, we derived four application scenarios in the context of continuous reverse engineering, which now can be realized by accessing a DArT using the AIQL.
Utilizing these future applications will decrease the maintenance caused by architectural drift.
In a user study with eight experts, they agreed that the AIQL is easy to understand and use. They further confirmed that AIQL queries enable drilling down into the architecture information in the DArT.

\section{The Digital Architecture Twin (DArT)}
\begin{figure*}[tbp]
    \centerline{\includegraphics[width=\linewidth]{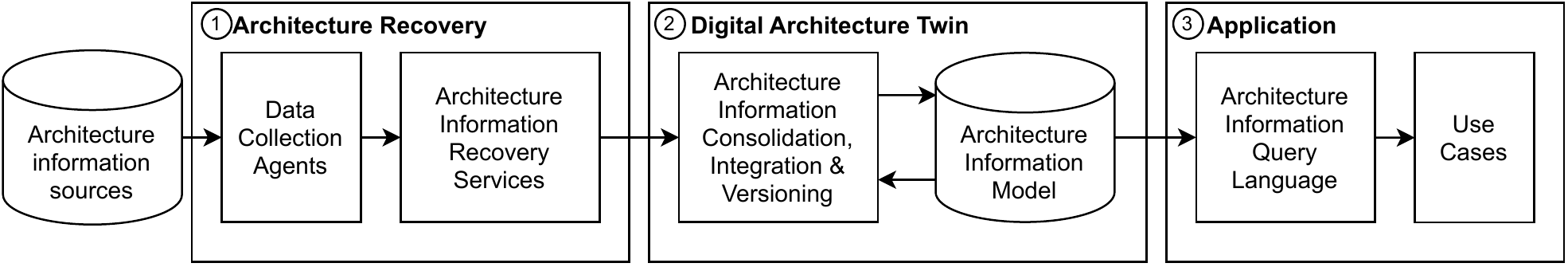}}
    \caption{Architecture Information Processing using the DArT~\cite{Jordan2022}.}
    \label{fig:dartProcess}
\end{figure*}

A DArT~\cite{Jordan2022} is an up-to-date digital representation of a software system. 
The DArT automatically recovers and integrates architecture information of multiple views and versions.
Fig.~\ref{fig:dartProcess} displays how the DArT is filled with information.
First, in an Architecture Recovery step (\raisebox{.5pt}{\textcircled{\raisebox{-.9pt} {1}}}), architecture data is continuously gathered from heterogeneous sources by Data Collection Agents.
Architecture Information Recovery Services apply different architecture recovery approaches on the gathered data to recover various architecture information.
The recovered architecture information is then integrated into the DArT as a virtual representation of a software system's architecture (step \raisebox{.5pt}{\textcircled{\raisebox{-.9pt} {2}}}). 
The architecture information is versioned and stored in a concise, uniﬁed and persistent Architecture Information Model.
In step \raisebox{.5pt}{\textcircled{\raisebox{-.9pt} {3}}} stakeholders want to access tailored parts of this information to fulfill specific tasks.
For this, an easy-to-use language is required to make the information accessible to various kinds of stakeholders, such as developers or enterprise architects.
\fontfamily{fontenc}\selectfont

\section{Application Scenarios}
\label{sec:ApplicationScenarios}
Frequent challenges in the industry are understanding and evolving software systems. 
This entails reusing of beneficial design patterns and the harmonization of software systems.
Another challenge is to ensure alignment of the implemented software system with the initially planned architecture.
This implies making architecture decisions for existing and new software systems using tailored and up-to-date architecture information, which supports the design decision process. 
We identified four application scenarios addressing these challenges, which define the requirements for the AIQL.
The application scenarios are refined into concrete use cases (\textsf{\textbf{UC}}s), which will be used later to evaluate the language.

\paragraph{Sce. 1: Pattern and style identification for continuous analysis}
The identification of reusable patterns and the exclusion of anti-patterns in software systems is beneficial for a high-quality maintainable software architecture.
Therefore the language should support the exploitation of patterns, anti-patterns, and design decisions, 
by enabling querying of structural and architecture information.
For example, a stakeholder may want to get a list of design patterns 
to capture them in a design pattern 
catalog (\textsf{\textbf{UC 1}}).

\paragraph{Sce. 2: Tailored documentation} 
Analyzing an architecture (or the available architecture information) is a necessity to make design decisions. 
To ease this, the language should support tailored architecture documentation in the form of different views for specific stakeholders.
Thus, querying for different views from an architecture information base needs to be supported.
The retrieved information may be used to strengthen the understanding of the system by visualizing it.
Examples for this application scenario are an enterprise architect, who wants to list all interfaces to another specific sub-system~(\textsf{\textbf{UC 2.1}}), or a product owner, who wants to create an overall structural diagram~(\textsf{\textbf{UC 2.2}}), or a developer, who wants to create a class diagram of a system component~(\textsf{\textbf{UC 2.3}}).
         
\paragraph{Sce. 3: Automated continuous compliance checking} 
The divergence of the planned architectural information from the implemented system can be detected at an early stage using automated techniques.
A query language should support the continuous comparison of the planned system architecture 
with the actual system architecture. 
For instance, an architect may want to run automated compliance checking against guidelines, which requires access to 
the latest version of the system architecture (\textsf{\textbf{UC 3}}). 

\paragraph{Sce. 4: Recommendation system for architectural design decisions}
A recommendation system for architectural design decisions enables the reuse of good design decisions of existing systems.
Such a recommendation system uses a
knowledge base of previous design decisions
to propose architectural designs that may benefit another software system.
For this, relevant system properties as well as design decisions, have to be accessible via a query language (\textsf{\textbf{UC 4.1}}).
Furthermore, a system architect might want to browse the knowledge-base
using exploratory queries (\textsf{\textbf{UC 4.2}}).

\paragraph{Language requirements} 
To be applicable to the above application scenarios, the architectural language to be developed has to provide information from various views.
Also, specific versions of a system should be queryable.
Further, the syntax of the language should be easy to use and understandable for a variety of stakeholders.
The architecture query language should capture the semantics of the software architecture domain and use existing notations that stakeholders are familiar with.
The output of a query should be presented in a usable and human-readable output format.
The output format should also support automation for further analysis and processing. 

\section{The Architecture Information Query Language (AIQL)}
\begin{figure}[tbp]
\sffamily
\centering
\begin{subfigure}[t]{0.29\textwidth}
    \centering
    \includegraphics[width=0.8\textwidth]{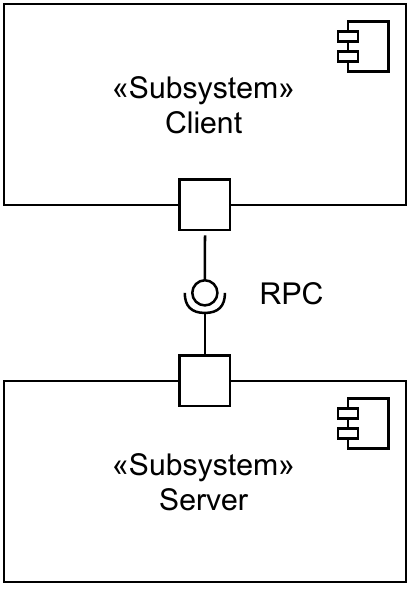}
    \caption{}
    \label{fig:clientServerComp}
\end{subfigure}
\begin{subfigure}[t]{0.7\textwidth}
    \includegraphics[width=\textwidth]{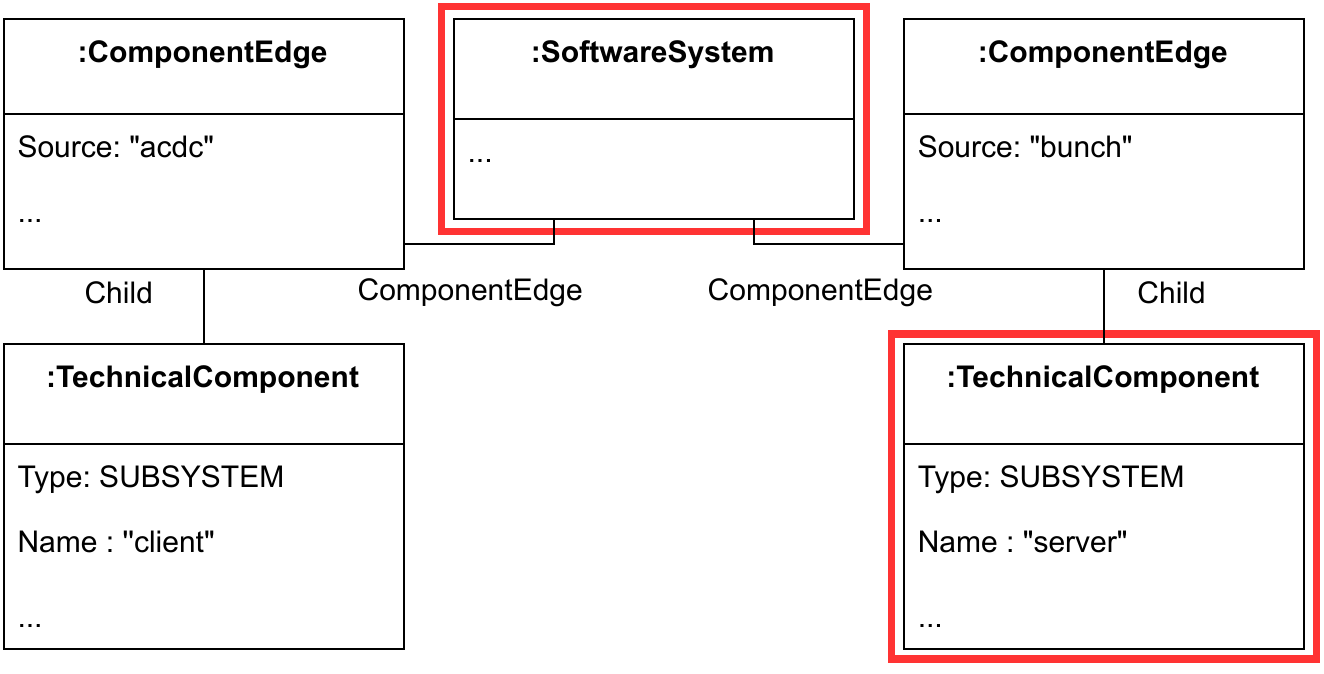}
    \caption{}
    \label{fig:dartInstance}
\end{subfigure}
\caption{Unified Modeling Language (UML) component diagram of a client-server example system (a). The UML object diagram shows the system recovered in the DArT (b). Highlighted are the objects that the AIQL query in Fig.~\ref{lst:exampleQuery} returns for this example system.}
\end{figure}

We designed the AIQL to satisfy the identified requirements.
In this section, we demonstrate the AIQL's main language features by an example.\footnote{Detailed description of the language's syntax in Extended Backus–Naur Form (EBNF) and its semantics are provided in appendix \ref{app:syntax} and \ref{app:semantics} respectively.}
Our running example consists of a client and a server component connected via Remote Procedure Call (RPC) (see Fig.~\ref{fig:clientServerComp}).
After the architecture recovery step, this system is represented in the DArT's technical component view as depicted in Fig.~\ref{fig:dartInstance}.
For each component from the running example, a corresponding \textsf{\textit{TechnicalComponent}} was created in the DArT. Further, a \textsf{\textit{SoftwareSystem}} component, representing the complete system, was created.

\begin{figure}[tb]
\begin{lstlisting}[
        morekeywords={MODEL, VERSION, LAST, LIST, RESTRICTIONS, OUTPUT, TechnicalComponent,
        Name, Source, Children, EXISTS, SoftwareSystem, Child, ComponentEdge, OR},
        keywordstyle=\sffamily\bfseries,
        frame=shadowbox,
        linewidth=\textwidth,
        xleftmargin=0.1\textwidth,
        xrightmargin=0.1\textwidth,
        numbers=left,
        stepnumber=1,
        escapechar=|]
MODEL "system.model"; |\label{line:model}|
VERSION LAST; |\label{line:version}|

LIST TechnicalComponent serverComponent RESTRICTIONS: |\label{line:list}|
    (Name 'server'); |\label{line:attr}|

LIST SoftwareSystem system RESTRICTIONS:
    (EXISTS Children serverComponent); |\label{line:ref}|

OUTPUT serverComponent;
OUTPUT system;
\end{lstlisting}
\caption{AIQL query example for client-server entities of the technical component view. The last version of the model \textsf{\textit{system.model}} is queried. Two query templates (\textsf{\textit{system}} and \textsf{\textit{serverComponent}}) are matched. Restrictions on the relation \textsf{\textit{Children}} and the attribute \textsf{\textit{Name}} are defined in the query templates.}
\label{lst:exampleQuery}
\end{figure}

\subsection{Language Design}
Imagine a software architect who wants to generate tailored documentation for server components (see scenario 2).
Fig.~\ref{lst:exampleQuery} shows an example AIQL query to retrieve server components from the DArT (for simplicity, we assume that such components could be identified by their name).
Using this example, we demonstrate the AIQL's main language features.

In general, an AIQL query consists of three parts: header, body, and output.
The header contains information for the whole query so that the AIQL is complete in the sense that no information has to be provided externally.
In the header (lines~\ref{line:model}-\ref{line:version}), the location of the \textsf{\textbf{MODEL}} from which the architecture information will be queried, and the \textsf{\textbf{VERSION}}(s) of the model are defined.
For the version, static keywords for the \textsf{\textbf{FIRST}} and \textsf{\textbf{LAST}} versions are provided, and multiple versions can be accessed through expressions.
Versions are required, e.g., for \textsf{\textbf{UC3}}, and are part of the header to ensure consistency in the query results.

In the query body, query templates for elements of the architecture information model can be defined, which a stakeholder either wants to output or use to restrict further in the query.
A query template is declared using the \textsf{\textbf{LIST}} directive (line~\ref{line:list}) followed by the unique name of the type of the element, that should be queried (e.g. \textsf{\textit{TechnicalComponent}}), and a user-defined identifier (e.g. \textsf{\textit{serverComponent}}).
\footnote{We refrained from reusing the Structured Query Language's (SQL) syntax (\textsf{\textbf{SELECT}} instead of \textsf{\textbf{LIST}}), as we want to highlight the fundamental differences in the underlying semantics:
instead of selecting columns from a table, in AIQL query templates result in a list of matched elements of a certain type.}
A query template matches the model elements satisfying the specified \textsf{\textbf{RESTRICTIONS}}.
Multiple restrictions can be defined in one query and are implicit conjunctions except if the \textsf{\textbf{OR}} keyword is used.
Restrictions can be used on element attributes (e.g. \textsf{\textit{Name}} in line~\ref{line:attr}) for which a static expression language for primitive types is provided.
For our example the String attribute \textsf{\textit{Name}} has to match the String \textsf{\textit{`server'}}.

The composition of query templates is a key element of the AIQL, that strongly encourages reuse.
References to other model elements can be restricted through a combination of 
a quantifier (such as \textsf{\textbf{EXISTS}} or \textsf{\textbf{FORALL}}), the name of the reference, and the identifier of another query template.
Quantifiers are only required for one-to-many or many-to-many references.
Entities of different views may be connected through relations and, thus, a stakeholder's queries can span multiple views.
In line~\ref{line:ref}, the one-to-many reference \textsf{\textit{Children}} from \textsf{\textit{SoftwareSystem}} to \textsf{\textit{TechnicalComponent}} is restricted by the \textsf{\textbf{EXISTS}} quantifier and the \textsf{\textit{serverComponent}} template.
Only software systems that have a server component are matched.
The reference \textsf{\textit{Children}} in this case is a shortcut defined in the AIQL as the DArT in this example contains \textsf{\textit{ComponentEdges}} (see Fig.~\ref{fig:dartInstance}) which are resolved internally.
Shortcuts allow more concise queries and can be user-defined.

Finally, architecture model elements can be output using multiple \textsf{\textbf{OUTPUT}} directives with the identifier of the query template to be output. 
Output directives support tailoring of the output format (e.g., to only output specific attributes). The query in Fig.~\ref{lst:exampleQuery} returns all software systems that have a server component and all server components contained in the latest version of the model. For the example, the model elements returned by this query are highlighted in Fig.~\ref{fig:dartInstance}. 

\subsection{Implementation}
To show the real-world applicability of the AIQL, we implemented a prototype for querying the DArT. 
Leveraging Model Driven Software Development (MDSD) techniques, the DArT's meta-model is a central artifact ``considered equal to code''~\cite{Stahl2006}
that incorporates domain-specific abstractions, in this case of the software architecture domain.
A concrete DArT instance has a concrete architecture information model that is compliant with the meta-model and filled with information of a concrete software system.

The concrete syntax of the AIQL is developed to match the DArT's meta-model using MDSD techniques and, thus, easily extensible\footnote{Due to confidentiality reasons as required by the industry partner, the source code cannot be provided, as the concrete language highly depends on the DArT's meta-model.}.
The available query templates and restrictions (their attributes and references) are derived directly from the DArT's meta-model.
The available expression language is static and applied according to the attribute types (e.g., in AIQL queries regular expressions can be used for \textsf{\textit{String}} attributes in the meta-model).
Additionally, users can define concise shortcuts which are accounted for in the AIQL's interpreter.

The implementation builds upon the Epsilon Object Language~\cite{Kolovos2006} and the Xtext language workbench~\cite{Eysholdt2010}.
An editor for manual use cases is provided as well as a query builder Application Programming Interface (API) for automation.
Executing queries via the API returns lists of Java objects,
while using the provided graphical editor and interpreter the results are marshaled into JavaScript Object Notation (JSON) format.
Fig.~\ref{lst:JSON} displays the JSON returned for the example query on the running example.
Leveraging code generation through Xtext and a model-to-text transformation, 
 the implementation can adapt to changes in the model with low effort,
  resulting in higher maintainability.

\lstdefinestyle{json}
{
  string=[s]{"}{"},
  stringstyle=\bfseries,
  comment=[l]{:},
  commentstyle=\color{black},
}

\begin{figure}[tb]
    \begin{lstlisting}[style=json,
        frame=shadowbox,
        linewidth=\textwidth,
        xleftmargin=0.2\textwidth,
        xrightmargin=0.2\textwidth,]
[      
    [ {
        "type" : "SoftwareSystem",
        ...
    } ], [ {
        "type" : "TechnicalComponent",
        "Type" : "SUBSYSTEM",
        "Name" : "server",
        "Version": ...,
        "ComponentEdge": [],
        ...
    } ]
]
    \end{lstlisting}
    \caption{Returned JSON for query of \textbf{UC2.2} executed on the example system.}
    \label{lst:JSON}
\end{figure} 

\section{User study} 
\label{sec:userstudy}

We performed a user study at an industry partner to further evaluate the AIQL's usability. The eight participants (Enterprise-, Solution-, and Domain Architects, IT Project managers, and developers) ranged from 7 to 21 years of work experience.
In the user study, we asked participants about the comprehensibility of the AIQL (produced output and language itself), the usability of the output and the complexity of example queries. In the user study we presented the AIQL in-depth (i.e., theoretical and practical) to the participants and allowed them to use the AIQL.
Furthermore, we showed participants the application scenarios and relevant example queries and asked them to assess the value of the AIQL for the application scenarios 1, 2 and 4.

\subsection{Research Questions}
We identified four research questions to evaluate the AIQL in terms of understanding, applicability and complexity: 
\begin{description}
    \item[\textbf{RQ1}]{Is the AIQL easy to comprehend and to use for stakeholders?}
    \item[\textbf{RQ2}]{Is the output of the AIQL comprehensible and useful to stakeholders?}
    \item[\textbf{RQ3}]{How complex are (nested) AIQL queries to use for stakeholders?}
    \item[\textbf{RQ4}]{How valuable is the AIQL for the application scenarios?}
\end{description}

Using these main research questions, we designed our user study questions to ask the participants if the output of the AIQL is understandable and if the output is useful to them. 
We further questioned participants if the syntax of the AIQL is easy to understand. We also asked them for valuable feedback regarding the power of the AIQL.

\subsection{Results and Discussion}

\subsubsection{RQ1}
All participants stated that the AIQL is easy to understand. In comparison to other languages (e.g. the Structured Query Language (SQL)), the AIQL was deemed equally easy. 62.5\% of participants found the language intuitive and the keywords meaningful. 
37.5\%\ of participants mentioned difficulties using the AIQL, due to the short timeframe of the user study, they were not able to memorize all keywords of the AIQL. More time to familiarize themselves with the AIQL as well as with the DArT was needed. 
Three participants (solution and enterprise architects) preferred the accessible language design, whereas one participant did not immediately grasp the composition of sub-queries.
It was also noted that the comprehensibility is limited for non-technical experts and that the language may be hard to understand for business users.
Detailed responses by the participants mentioned missing an ``Aggregate''-keyword to automatically filter the highest-level sub-systems. The current editor support was praised to ease getting started with the AIQL.

\subsubsection{RQ2}
All participants stated that the AIQL is useful for extracting architecture information about a software system, but it is restricted by the underlying data model.
50\% of participants noted that the AIQL, if not used to full power is not necessarily more useful than a static view, but the AIQL is more useful when used in an exploratory approach.
It was debated that the AIQL is more useful, when adding information about frequency of class calls and when the stakeholder knows the system to some degree. A supplementing static/dynamic view would be help for an initial overview of the software system.
37,5\% of participants found the AIQL more useful for drilling down into the architecture information. 
It was noted that the AIQL does not provide more information than a static visualisation as it only displays recovered architecture information.
Participants mentioned that the output format (JSON) is hard for humans to read, but useful for further use and automation.

\subsubsection{RQ3}
Participants stated that the complexity of shown example queries varied from trivial to complex.
An easy example query was perceived as very simple. An example query in the complexity of the running example was still perceived as easy. Participants mentioned that linking queries using the sub-query templates was perceived as doable.
A complex query using four nested sub-queries was deemed as complex, leading participants to express concerns about the complexity of queries containing more nested sub-queries. Some participants also criticized the lack of self-contained queries (i.e. no sub-query composition).

\subsubsection{RQ4}
50\% of participants stated that the AIQL is useful to extract tailored architecture information. 
Depending on the participant's background, some find the AIQL to be potentially more useful.
Developers or solution architects assessed the AIQL as more useful for extracting tailored architecture information for specific use cases than enterprise architects.
The participants believed that the AIQL can be helpful in understanding the source code, from an unknown software system.
All participants stated the use of a recommendation system or wizard-based queries for finding similar software systems based on requirements and architecture is very interesting and "incredibly useful".
They also discussed the potential benefits of the AIQL by understanding functional requirements and how it can be useful for finding software components and libraries used for these requirements.
One participant stated that the use of AIQL in combination of the DArT is potentially helpful in finding an appropriate point of contact for further insights.
The participants mentioned two main aspects which the language was lacking: comprehensibility of the output and aggregation of architectural information. Future work is needed to provide other output formats or to provide aggregation in the AIQL. 
Whereas the former depends on the output format and requires changes to the AIQL's interpreter, the latter requires enhancement of the underlying DArT/data model, which has to provide the necessary information, so that the AIQL is able to aggregate architecture information. 

\section{Empirical Evaluation}
The AIQL is evaluated quantitatively in this section regarding its applicability in the defined application scenarios (cf.~\ref{sec:ApplicationScenarios}), its usability, and its scalability.

\subsection{Research Questions and Methodology}
\begin{figure}[tb]
    \sffamily
    \centerline{\includegraphics[width=\textwidth]{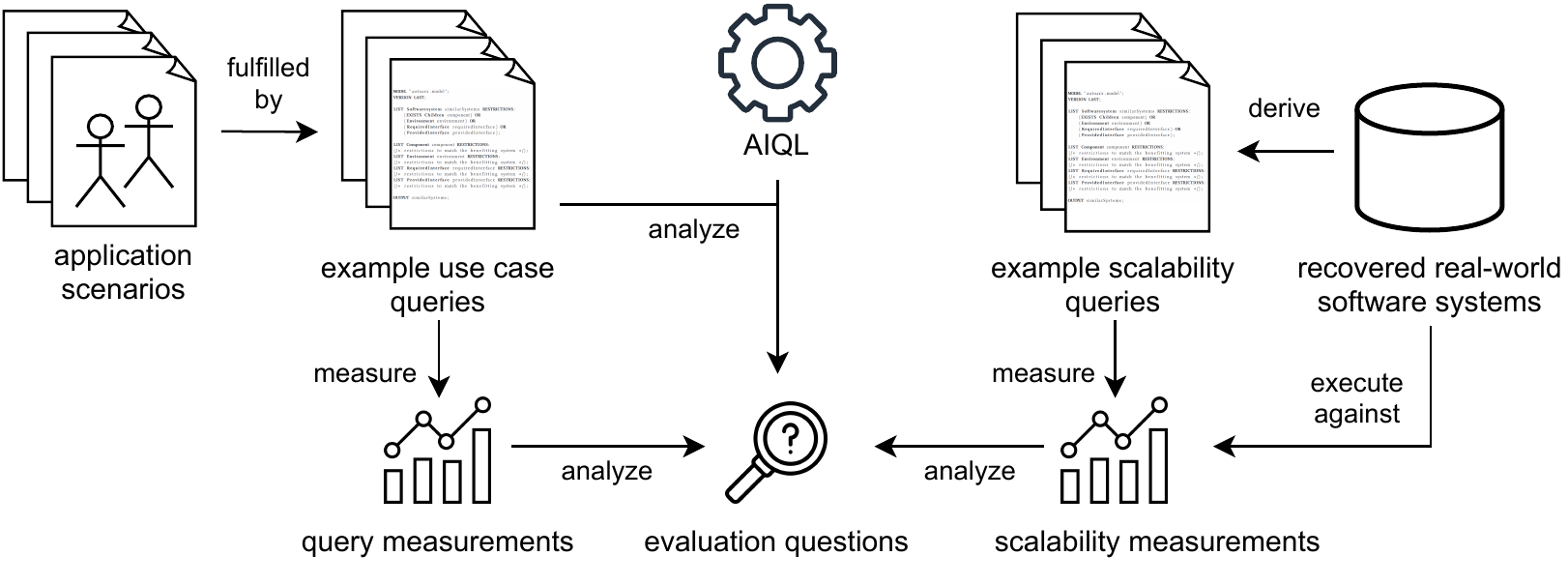}}
    \caption{The quantitative evaluation process of the AIQL. AIQL queries for the application scenarios are defined and their characteristics are measured. The measurements are analyzed to answer the research questions.}
    \label{fig:evlProcess}
\end{figure}

We derived the following three research questions and conducted the evaluation process depicted in Fig.~\ref{fig:evlProcess}.

\begin{description}
    \item[\textbf{RQ5}]{Can all defined application scenarios be realized appropriately using the AIQL?}
    \item[\textbf{RQ6}]{How many inputs by a user are necessary to formulate AIQL queries?}
    \item[\textbf{RQ7}]{Does the AIQL scale for real-world systems?}
\end{description}

\textbf{RQ5} investigates if the AIQL contains all necessary capabilities and checks if it is possible to map the requirements of the application scenarios to the syntax of the AIQL.
\textbf{RQ6} is concerned with usability in a quantitative fashion (for a qualitative user study see Sect. ~\ref{sec:userstudy}). Having fewer mandatory inputs increases the likelihood of acceptance by the user. In this case, we envision the AIQL as a tool to drill into the existing architecture information which is a manual process. A stakeholder should be supported with the manual process as much as possible.
\textbf{RQ7} is concerned with the scalability of the AIQL in terms of query execution times and asks if the AIQL scales for large systems. This is important as large systems can contain many different architecture elements and providing architecture information to the stakeholder has to be in a timely manner.

\subsection{Results and Discussion}
\subsubsection{RQ5} We defined queries that retrieve the information required to realize the use cases corresponding to application scenarios.
Even though the diverse use cases required queries to retrieve information from multiple architectural views and a range of complexities, a single AIQL example query was sufficient for each use case. We implemented a recommendation system using AIQL queries to identify similar software systems based on (non)-functional requirements. Also possible is the support of the tailored documentation, as the AIQL can be used to gather the needed architecture information before processing them in a manual process to create the specific diagrams.
The output could be customized and was presented in JSON format (Fig.~\ref{lst:JSON}). Hence, we can answer the RQ positively that the AIQL is able to handle the defined application scenarios.

\begin{table}[tb]
    \scriptsize
    \caption{Measurements of query characteristics for example AIQL queries.}
    \begin{center}
    \begin{tabular}{|l|x{0.7cm}|x{0.7cm}|x{0.7cm}|x{0.7cm}|x{0.7cm}|x{0.7cm}|x{0.7cm}|c|}
        \hline
        \textbf{Use case id} & \textbf{1} & \textbf{2.1} & \textbf{2.2} & \textbf{2.3} & \textbf{3} & \textbf{4.1} & \textbf{4.2} & \textbf{Avg.} \\\hline
        \textbf{Number of queries} & 1 & 1 & 1 & 1 & 1 & 1 & 1 & 1 \\\hline
        \textbf{Number of query outputs} & 1 & 2 & 1 & 1 & 3 & 1 & 2 & 1.57 \\\hline
        \textbf{Number of query characters} & 53 & 112 & 58 & 154 & 103 & 297 & 165 & 134.57 \\\hline
        \textbf{Min. number of input keystrokes} & 33 & 65 & 33 & 87 & 68 & 148 & 119 & 79 \\\hline
        \textbf{Number of keywords in total} & 6 & 11 & 6 & 15 & 12 & 27 & 17 & 13.43 \\\hline
        \textbf{Number of unique keywords} & 6 & 9 & 6 & 12 & 8 & 14 & 13 & 9.71 \\\hline
    \end{tabular}
    \end{center}
    \label{tab:queryMeasurement}
\end{table}

\subsubsection{RQ6} Table~\ref{tab:queryMeasurement} depicts the quantitative measurements conducted on the use case example queries.
The number of queries for each use case was counted by the number of interpreter executions required in the editor.
For the number of query characters (the query length), we counted dynamic-sized elements such as identifiers and strings as constant 1.
Additionally, only syntactically relevant white spaces (e.g. between keywords) were counted 
and comments were ignored.
The minimal number of input keystrokes was measured using the provided editor utilities of the realization,
where key combinations (e.g. CTRL + enter) were counted as multiple keystrokes.
Elements, white spaces, and comments were counted similarly to the query length.
For the number of query outputs, the output directives of each query were counted.
(Unique) keywords (that are highlighted as bold in the queries) were counted for the last two measurements.

From these measurements, we conclude the following regarding the AIQL's usability:
Using multiple output directives, only one query is needed for every use case.
The query length varies depending on the query's complexity, averaging 135 characters for the examples.
Making full use of provided language smarts (auto-complete and proposals),  significantly reduces the number of input keystrokes in the examples: only 79 keystrokes on average instead of 135 when typing every character.
The AIQL supports 18 syntactic constructs that the user has to remember.
The total number of keywords used increases depending on the concrete underlying architecture information model.
For the example queries, the average query made use of 13 keywords in total and 10 unique keywords.
By using language smarts provided in the AIQL editor, the user does not have to remember the keywords.

\subsubsection{RQ7} To evaluate the AIQL's scalability, open-source real-world software systems are transformed into the DArT.
We used the mid-size system \textbf{fastjson} (44 kLOC) and the two large-size systems \textbf{Apache Hadoop} (183 kLOC) and \textbf{Apache Struts2} (161 kLOC).
These systems were chosen because they differ in size (mid to large) as well as structuredness (grown to well structured). 
We defined seven AIQL queries\footnote{The queries are provided in appendix \ref{app:scalQueries}.} 
that address only the structural view, as this information is recovered directly from the code base and assumed to be the largest view in the DArT.
The queries were selected to differ only in specific language elements, such as restrictions and template nesting, to show how these elements influence runtime.
The AIQL queries are executed for the recovered systems and their run times are measured and analyzed.
For each query, its description and characteristics are depicted on the left side of Table~\ref{tab:scalability}.
The measured execution time of the query (including interpretation) and the output marshaling times are depicted on the right side of Table~\ref{tab:scalability}.
The measurements were conducted on a Windows Desktop Computer with Intel(R) Core(TM) i7-6700K CPU @ 4.00GHz and 16 GB of RAM.
Fig.~\ref{fig:scalability} visualizes the overall query execution and marshaling time.

\begin{table}[tb]
    \sffamily
    \scriptsize
    \caption{AIQL queries and scalability measurements using real-world systems.}
    \begin{center}
    \begin{tabular}{|x{0.4cm}|l|x{0.6cm}|c|c|c||D{,}{ + }{4.4}|D{,}{ + }{4.4}|D{,}{ + }{4.4}|}
        \hline
         & & \multicolumn{4}{c||}{\textbf{Number of}} & \multicolumn{3}{c|}{\textbf{Execution + marshaling}}\\
         & & \textbf{out-} & \textbf{tem-}  & \textbf{refer-} & \textbf{expres-}  & \multicolumn{3}{c|}{\textbf{time in \textit{ms}}} \\
         \textbf{Id} & \textbf{Description} & \textbf{puts} & \textbf{plates} & \textbf{ences} & \textbf{sions} & \multicolumn{1}{c|}{\textbf{fastjson}} & \multicolumn{1}{c|}{\textbf{Hadoop}} & \multicolumn{1}{c|}{\textbf{Struts2}}\\
        \hline
        \textbf{1} & technical components        & 1 & 1 & 0 & 0 &   52 , 16 &  143 , 53 &  208 , 55 \\\hline
        \textbf{2} & classes                     & 1 & 1 & 0 & 1 &   54 , 7 &  165 , 26 &  210 , 18 \\\hline
        \textbf{3} & classes named handler       & 1 & 1 & 0 & 2 &   61 , <1  &  184 , 2 &  238 , 1 \\\hline
        \textbf{4} & edges to handler classes    & 1 & 2 & 1 & 2 &   86 , <1 &  292 , 1 &  522 , 2 \\\hline
        \textbf{5} & edges AND handler classes   & 2 & 2 & 0 & 2 &  132 , 23&  397 , 32 &  463 , 47 \\\hline
        \textbf{6} & parents of handler classes  & 1 & 3 & 2 & 2 &  147 , 3 &  657 , 31 &  934 , 35 \\\hline
        \textbf{7} & edges AND handler classes   & 3 & 3 & 0 & 2 &  220 , 47 &  466 , 82 &  568 , 87 \\
        & AND components & & & & & & & \\ \hline
    \end{tabular}
    \end{center}
    \label{tab:scalability}
\end{table}

\begin{figure}[tb]
\sffamily
\centering
\begin{tikzpicture}
\begin{axis}[
    x tick label style={
        /pgf/number 
        format/1000 sep=},
    ylabel=Execution time in \textit{ms},
    enlargelimits=0.15,
    legend style={
        at={(0.2,0.99)},
        anchor=north},
    ybar,
    bar width=7pt,
    ymajorgrids,
    scaled ticks=false,
    xtick=\empty,
    xmin=0,
    xmax=6,
    xticklabel=\empty,
    extra x ticks={0.5,3,5.5},
    extra x tick labels={fastjson, Apache Hadoop, Apache Struts2},
    extra x tick style={major tick length=0pt},
    width=\textwidth,
    height=.3\textheight
]
\addplot[fill=PineGreen!25]
	coordinates {(0.5, 52+16) (3, 143+53) (5.5, 208+55)};
\addplot[fill=PineGreen!50]
	coordinates {(0.5, 54+7) (3, 165+26) (5.5, 210+18)};
\addplot[fill=PineGreen!75]
	coordinates {(0.5, 61+0) (3, 184+2) (5.5, 238+1)};
\addplot[preaction={fill, Purple!50}, pattern=crosshatch]
	coordinates {(0.5, 86+0) (3, 292+1) (5.5, 522+2)};
\addplot[fill=Purple!70]
	coordinates {(0.5, 132+23) (3, 397+32) (5.5, 463+47)};
\addplot[preaction={fill, RoyalBlue!50}, pattern=crosshatch]
	coordinates {(0.5, 147+3) (3, 657+31) (5.5, 934+35)};
\addplot[fill=RoyalBlue!70]
	coordinates {(0.5, 220+47) (3, 466+82) (5.5, 568+87)};
\legend{
\scriptsize Query 1 (1 Template),
\scriptsize Query 2 (1 Template), \scriptsize Query 3 (1 Template),
\scriptsize Query 4 (2 Templates), 
\scriptsize Query 5 (2 Templates), 
\scriptsize Query 6 (3 Templates), 
\scriptsize Query 7 (3 Templates)}
\end{axis}
\end{tikzpicture}
\caption{Overall AIQL query execution and marshaling time in \textit{ms} for each system and query (see Table~\ref{tab:scalability}). The AIQL scales well for real-world systems. But, as expected, introducing new query templates increases query time. Restrictions on references (crosshatched queries) greatly impact query time for larger systems.}
\label{fig:scalability}
\end{figure}

From the results, we can see that the queries are executed in a reasonable time for real-world systems, below 1 $second$ for the example queries.
A small number of simple restrictions does not affect query execution time significantly for all considered systems (\textit{queries 1 to 3}).
Introducing reference restrictions (\textit{crosshatched queries}) significantly increases execution time, roughly doubling for large systems.
This implies that nested reference restrictions may become a bottleneck for large systems.
Simply outputting all templates instead of using reference restrictions (\textit{not crosshatched vs crosshatched queries} with the same amount of templates) perform better for large systems and queries and may be an alternative.

\subsection{Threats to Validity}

\subsubsection{Internal Threats}
Scalability measurements of the queries were executed and measured on a standard computer, thus its operating system's scheduler influenced the measured time. To address an introduced bias, the queries were executed 5 times and the average was taken, nevertheless, replicability of the exact same measurements can not be assured.
These measurements were conducted using our own implementation of the AIQL's interpreter and API, which we tested for small samples, but cannot assure that no errors occurred.

\subsubsection{External Threats}
The evaluation is based on predefined application scenarios and use cases that
should be enabled through the AIQL.
Thus, there is the possibility that these scenarios and use cases are not representative. 
Furthermore, the defined use case queries may also not represent
actual user queries correctly, e.g., many concrete restrictions may be larger,
and the queries may be biased, as they were formulated by the languages designers.
Regarding this bias, another user study could mitigate such uncertainties in the future. 
For the scalability evaluation, we defined queries to assess the influence 
of language constructs towards run time performance in the context of real-world systems.
We carefully selected queries to isolate the effects of certain language constructs,
but still, the evaluation may be biased by the selection of queries as well as by the example systems. 

\section{Related Work}
\label{sec:Related work}
Querying of software systems is frequently used in the context of Software Architecture Recovery (SAR).
The following approaches stated in~\cite{Ducasse2009} are closest to our approach, as they use queries to present architectural information:
\textit{ARM} by Guo et al. (1999)~\cite{Guo1999}, 
\textit{Alborz} by Sartipi (2001)~\cite{Sartipi2001} and 
\textit{Revealer} by Pinzger et al. (2002)~\cite{Pinzger2002Revealer} focus on architectural patterns.
\textit{ArchView} also by Pinzger (2005)~\cite{Pinzger2005Thesis} provides architectural views.
Relational queries are the most common query technologies and are used by tools such as ArchView, ARM, and Kazman and Carri{\`{e}}re's \textit{Dali} (1999)~\cite{Kazman1999}~\cite{Ducasse2009}.
Dali and ArchView directly use SQL on an architectural database~\cite{Ducasse2009}. Both approaches persist the recovered architecture information in a SQL database, which is queried to retrieve the recovered architecture information, however both approaches do not correlate the architecture information in the database with each other.
ARM allows the definition of pattern descriptions and rules in a commonly used format (Rigi Standard Format) that are translated into pattern queries exploitable in Dali via SQL~\cite{Guo1999}.
Lexical queries in Revealer can be defined using a simple pattern language based on regular expressions~\cite{Ducasse2009}.
All these tools except Alborz use generic query mechanisms that are not tailored to the domain of software architecture.
However, the architecture information in the database is not correlated with other architecture information.
Our proposed AIQL is a Domain-Specific Language (DSL) instead of a general-purpose query language.
DSLs leverage the restriction to a particular problem domain - the software architecture domain in our case - to gain expressive power by providing domain-tailored notations and abstraction.

\paragraph{Architectural Query Languages (AQLs)} In contrast, AQLs are tailored for architecture information retrieval.
The term AQL was proposed by Sartipi in the context of an interactive reverse engineering environment for pattern-based
recovery of software architecture information~\cite{Sartipi2003Thesis}, that is implemented in the Alborz toolkit.
The goal of this environment was to address the lack of a 
reflective and uniform model for pattern-based software architecture recovery. 
Achitecture information is incrementally recovered
by defining and refining patterns defined in a proposed AQL, which are matched against a database of architectural information,
 that is generated by the environment in an offline process.
A query in the language consists of abstract connectors and components inspired by ADLs.
The language itself supports entities on file and function level, typed import, export, and containment relations,
and constraints on the composition and interconnection size and type between subsystems or modules.
Although Sartipi's AQL originates from an architecture recovery background,
the tight coupling of Sartipi's AQL to the manual recovery process using special syntactic elements makes it difficult to reuse aspects for a language that is not driven by the proposed process, whereas the AIQL is independent of the architecture recovery process.
Wang and Gupta (2004) defined the query language \textit{ADLQL}~\cite{Wang2004}.
They aim to enable information exchange between ADLs, which are not expressive enough by themselves.
Thus, a framework enabling integration of ADL is introduced.
ADLQL allows the definition of queries that can be executed against predicates, representing architectural information, using denotational semantics. 
As ADLQL is inspired by SQL, common language constructs are reused and an OCL-like expression syntax is provided.
ADLQL is limited to ADLs, which have to be integrated and do not 
account for other relevant artifacts, such as a system's source code, whereas the AIQL is capable of handling source code.
Michalik (2010) proposed a framework supporting tailored architecture views
of a software product line's architectures using an AQL in~\cite{Michalik2010}.
It was realized as an extensible API querying an architecture knowledge repository and not as a language~\cite{Weyns2011}.
Monroy et al. (2021) introduced a query mechanism for recovered architecture models using natural language (Spanish) in the \textit{ARCo} framework~\cite{Monroy2021}.
An advantage of the natural language is its richness benefiting inexperienced stakeholders, for example, students, 
but on the other side, it is ambiguous~\cite{Monroy2021, Taylor2010}.
The use of more semantically narrow and formal languages is more suitable,
as we expect our stakeholders to have a decent amount of domain knowledge.

\paragraph{Detailed Comparison} Tab.~\ref{tab:languageComparison} compares the proposed AIQL to the state-of-the-art languages AQL and ADLQL.
All three languages have different purposes leading to differences in language design.
The AQL incorporates an iterative architecture recovery process through query refinement, thus the final output is the refined AQL query.
As the AQL is limited in the supported views and constrained to a specific recovery process, it is too restrictive to use in another context.
For extensive architecture data models, some information may be irretrievable if there is no suitable view to query it.
ADLQL and AIQL focus on querying architecture information, so the languages are process agnostic and support the reuse of queries for different target systems.
The AIQL provides a typed expression language in contrast to ADLQL's OCL-like expressions.
ADLQL focuses on ADL documents such that the language is extensible by adding support for further ADLs and the output of a query is an ADL document.
This is inconvenient to stakeholders because for each view a suitable ADL has to be learned and used.
In contrast, the AIQL addresses the research gap of querying recovered architecture information and outputs lists of instances from the domain model.
The AIQL is extensible through MDSD techniques.

\begin{table}[tb]
    \scriptsize
    \sffamily
    \caption{Comparison of AQL, ADLQL, and AIQL.}
    \begin{center}
    \begin{tabular}{|l|c|c|c|}
        \hline
        & \textbf{AQL by Sartipi~\cite{Sartipi2003Thesis}} & \textbf{ADLQL by Wang~\cite{Wang2007Thesis}} & \textbf{AIQL} \\\hline
        \textbf{Purpose} & iterative architecture & query ADL  & query recovered \\
        & recovery & documents & architecture information \\\hline
        \textbf{Process agnostic} & no & yes & yes \\\hline
        \textbf{Query reusability} & no & yes & yes \\\hline
        \textbf{Extensible} & no & through ADLs & through meta-model \\\hline
        \textbf{Output format} & AQL & ADL & domain model instances \\\hline
        \textbf{Expressions} & no & OCL-like & typed expression language \\\hline
    \end{tabular}
    \end{center}
    \label{tab:languageComparison}
\end{table}

\section{Conclusion and Future Work}
\label{sec:Conclusion}
This work aimed to provide an architecture query language to access the information present in extensive architecture data models such as DArTs.
Based on application scenarios of such a language in the reverse engineering context,
we proposed the AIQL and demonstrated its language features in this paper.
The language encourages reuse through the composition of templates and supports
querying information of different views and levels of abstraction along
with different versions of a system's architecture over time.
In the empirical evaluation, we showed that the AIQL provides the required functionality to formulate queries for application scenarios in continuous reverse engineering and that the language scales for use with real-world software systems.
In a  user study, we showed that the AIQL is easy to understand, but also identified two aspects to improve our language: providing additional language constructs and providing other output formats.
The user study also highlighted the demand for a supporting static view.
Future work is concerned with the realization of a continuous reverse engineering application scenario using the AIQL and leveraging the API for automation purposes.

\bibliographystyle{splncs04}
\bibliography{bibliography}

\appendix
\newpage
\section{AIQL syntax in EBNF}
\label{app:syntax}

\begin{figure}[htb]
    \footnotesize
    \begin{framed}
        \begin{grammar}
            \input{aiqlMain.tex}
            
            \input{aiqlTemplate.tex}
            
            \input{aiqlOutput.tex}
        \end{grammar}
    \end{framed}
\caption{AIQL main grammar rules in EBNF.}
\label{gra:aiql}
\end{figure}

\clearpage
\begin{figure}[!htb]
    \footnotesize
    \begin{framed}
        \begin{grammar}
            \input{aiqlExpressions.tex}
            
            \input{aiqlRemainder.tex}
        \end{grammar}
    \end{framed}
\caption{AIQL expression grammar rules in EBNF.}
\label{gra:aiql}
\end{figure}

\newpage
\section{AIQL semantics}
\label{app:semantics}

Let $C$ be the set of all query-able classes defined in the domain model
with the set $R$ of all directed relations $r \in R: c_1 \rightarrow c_2$ where $ c_1, c_2 \in C$.
The model also includes a set of all attributes $A$ of query-able classes.
Now for each class $c \in C$ two sets containing only the attributes $A_c$ 
and references $R_c: c \rightarrow c_{ref}$ can be derived,
where the referenced class $c_{ref}$ can be any class 
so that $c_{ref} \in C$.
Also, $A_c \subset A$ and $R_c \subset R$ have to hold.

For a template $t$ the \textsf{\textit{typeName}} has to correspond to the name of a class $c$ in $C$.
Templates have a unique identifier assigned in the \textsf{\textbf{LIST}} directive, by which they can be referenced
in the \textsf{\textbf{OUTPUT}} directive and in references.
Attributes $A_t$ of the template $t$ that are restricted in the template's conjunctions 
have to be a subset of the available attributes in that class $A_c$, so that
$A_t \subset A_c$ and they are identified by their \textsf{\textit{attrName}}.
The same holds for referenced attributes by their \textsf{\textit{attrName}} in the \textsf{\textbf{OUTPUT}} and 
\textsf{\textbf{ORDER_BY}} directives, which have to correspond to the class of the referenced template.
Attribute expressions, e.g., for float and int attributes, should have correctly typed literals. 
Similar to attributes, references $R_t$ are identified by their \textsf{\textit{refName}}, and only outgoing relations from
the current class $R_c$ can be restricted on $t$ as $R_t \subset R_c$.
Quantifiers for references have to be provided if the maximal cardinality of the 
referenced element class $c_{ref}$ in the relation
is greater than one: $|c_{ref}| > 1$, thus multiple referenced elements are possible.
Otherwise, no quantifier should be provided.

Furthermore, we allow the definition of additional shortcut relations
$s \in S$ mapping classes $c_1 \rightarrow c_2$ where $c_1, c_2 \in C$ that have not already been defined
in $R$, so that $R \cap S = \emptyset$.
For quantification, the same constraints as for relations have to hold for shortcuts. 
These shortcuts can be used to enable shorter queries for common references over multiple edges.

\newpage
\section{Queries for scalability evaluation}
\label{app:scalQueries}

\begin{figure}[htb]
\begin{lstlisting}[
        morekeywords={MODEL, VERSION, LAST, LIST, RESTRICTIONS, OUTPUT, TechnicalComponent,
        Name, Source, Children, EXISTS, SoftwareSystem, Child, ComponentEdge, OR, Type, CLASS},
        keywordstyle=\sffamily\bfseries,
        frame=shadowbox,
        linewidth=\textwidth,
        xleftmargin=0.1\textwidth,
        xrightmargin=0.1\textwidth,
        stepnumber=1,
        escapechar=|]
MODEL "{corresponding model}";
VERSION LAST;

LIST TechnicalComponent comp;

OUTPUT comp;
\end{lstlisting}
\caption{AIQL query 1 for scalability evaluation: Output all technical components.}
\end{figure}

\begin{figure}[htb]
\begin{lstlisting}[
        morekeywords={MODEL, VERSION, LAST, LIST, RESTRICTIONS, OUTPUT, TechnicalComponent,
        Name, Source, Children, EXISTS, SoftwareSystem, Child, ComponentEdge, OR, Type, CLASS},
        keywordstyle=\sffamily\bfseries,
        frame=shadowbox,
        linewidth=\textwidth,
        xleftmargin=0.1\textwidth,
        xrightmargin=0.1\textwidth,
        stepnumber=1,
        escapechar=|]
MODEL "{corresponding model}";
VERSION LAST;

LIST TechnicalComponent comp RESTRICTIONS:
(
	Type CLASS
);

OUTPUT comp;
\end{lstlisting}
\caption{AIQL query 2 for scalability evaluation: Output all technical components that are classes.}
\end{figure}

\begin{figure}[htb]
\begin{lstlisting}[
        morekeywords={MODEL, VERSION, LAST, LIST, RESTRICTIONS, OUTPUT, TechnicalComponent,
        Name, Source, Children, EXISTS, SoftwareSystem, Child, ComponentEdge, OR, Type, CLASS},
        keywordstyle=\sffamily\bfseries,
        frame=shadowbox,
        linewidth=\textwidth,
        xleftmargin=0.1\textwidth,
        xrightmargin=0.1\textwidth,
        stepnumber=1,
        escapechar=|]
MODEL "{corresponding model}";
VERSION LAST;

LIST TechnicalComponent comp RESTRICTIONS:
(
	Type CLASS
	Name '.*Handler'
);

OUTPUT comp;
\end{lstlisting}
\caption{AIQL query 3 for scalability evaluation: Output all technical components that are classes and that are named 'Handler'.}
\end{figure}

\begin{figure}[htb]
\begin{lstlisting}[
        morekeywords={MODEL, VERSION, LAST, LIST, RESTRICTIONS, OUTPUT, TechnicalComponent,
        Name, Source, Children, EXISTS, SoftwareSystem, Child, ComponentEdge, OR, Type, CLASS},
        keywordstyle=\sffamily\bfseries,
        frame=shadowbox,
        linewidth=\textwidth,
        xleftmargin=0.1\textwidth,
        xrightmargin=0.1\textwidth,
        stepnumber=1,
        escapechar=|]
MODEL "{corresponding model}";
VERSION LAST;

LIST ComponentEdge edge RESTRICTIONS:
(	
	Child handler
);

LIST TechnicalComponent handler RESTRICTIONS:
(
	Type CLASS
	Name '.*Handler'
);

OUTPUT edge;
\end{lstlisting}
\caption{AIQL query 4 for scalability evaluation: Output all edges to handler technical components.}
\end{figure}

\begin{figure}[htb]
\begin{lstlisting}[
        morekeywords={MODEL, VERSION, LAST, LIST, RESTRICTIONS, OUTPUT, TechnicalComponent,
        Name, Source, Children, EXISTS, SoftwareSystem, Child, ComponentEdge, OR, Type, CLASS},
        keywordstyle=\sffamily\bfseries,
        frame=shadowbox,
        linewidth=\textwidth,
        xleftmargin=0.1\textwidth,
        xrightmargin=0.1\textwidth,
        stepnumber=1,
        escapechar=|]
MODEL "{corresponding model}";
VERSION LAST;

LIST ComponentEdge edge;

LIST TechnicalComponent handler RESTRICTIONS:
(
	Type CLASS
	Name '.*Handler'
);

OUTPUT handler;
OUTPUT edge;
\end{lstlisting}
\caption{AIQL query 5 for scalability evaluation: Output all edges and also output all handler technical components.}
\end{figure}

\begin{figure}[htb]
\begin{lstlisting}[
        morekeywords={MODEL, VERSION, LAST, LIST, RESTRICTIONS, OUTPUT, TechnicalComponent,
        Name, Source, Children, EXISTS, SoftwareSystem, Child, ComponentEdge, OR, Type, Component, CLASS},
        keywordstyle=\sffamily\bfseries,
        frame=shadowbox,
        linewidth=\textwidth,
        xleftmargin=0.1\textwidth,
        xrightmargin=0.1\textwidth,
        stepnumber=1,
        escapechar=|]
MODEL "{corresponding model}";
VERSION LAST;

LIST Component handlerParent RESTRICTIONS:
(
	EXISTS ComponentEdge edge
);

LIST ComponentEdge edge RESTRICTIONS:
(
	Child handler
);

LIST TechnicalComponent handler RESTRICTIONS:
(
	Type CLASS
	Name '.*Handler'
);

OUTPUT handlerParent;
\end{lstlisting}
\caption{AIQL query 6 for scalability evaluation: Output all Components that have a handler technical component child.}
\end{figure}

\begin{figure}[htb]
\begin{lstlisting}[
        morekeywords={MODEL, VERSION, LAST, LIST, RESTRICTIONS, OUTPUT, TechnicalComponent,
        Name, Source, Children, EXISTS, SoftwareSystem, Child, ComponentEdge, OR, Type, Component, CLASS},
        keywordstyle=\sffamily\bfseries,
        frame=shadowbox,
        linewidth=\textwidth,
        xleftmargin=0.1\textwidth,
        xrightmargin=0.1\textwidth,
        stepnumber=1,
        escapechar=|]
MODEL "{corresponding model}";
VERSION LAST;

LIST Component component;

LIST ComponentEdge edge;

LIST TechnicalComponent handler RESTRICTIONS:
(
	Type CLASS
	Name '.*Handler'
);

OUTPUT component;
OUTPUT edge;
OUTPUT handler;
\end{lstlisting}
\caption{AIQL query 7 for scalability evaluation: Output all components, edges, and handler technical components.}
\end{figure}

\end{document}